\documentstyle[12pt]{article}
\begin{document}
\title{From unpolarized to polarized quark distributions in the nucleon
\thanks{Work supported in part by the KBN-Grant 2-P302-143-06.}}
\author{Jan Bartelski\\
Institute of Theoretical Physics, Warsaw University,\\
Ho$\dot{z}$a 69, 00-681 Warsaw, Poland. \\ \\
\and
Stanis\l aw  Tatur \\ 
Nicolaus Copernicus Astronomical Center,\\ Polish Academy of Sciences,\\
Bartycka 18, 00-716 Warsaw, Poland.}
\date{}
\maketitle
\begin{abstract}
\noindent  Starting from Martin, Roberts and Stirling fit for unpolarized
deep inelastic structure functions and using the newest experimental data
on spin asymmetries we get a fit which provides polarized quark 
distributions. We analyze the behaviour of such functions near $x$ equal to 1. 
The first moments of these distributions are also discussed. Our fit prefers
combination of proton and neutron data versus proton-deuteron one.
\end{abstract}
\newpage
Recently the interest in the spin structure of the nucleon  has been renewed
due to new data taken in experiments at CERN \cite{d,p} and SLAC \cite{n}.
The polarized deep inelastic asymmetries for deuteron and proton (Spin Muon
Collaboration at CERN) as well as for neutron (at $^{3}He$ target in 
E 142 experiment at SLAC) 
were precisely measured also in small $x$ region. Together with an old SLAC
\cite{pold} and EMC \cite{emc} data for proton one has considerable amount
of information which can be used to study nucleon spin structure and 
particularly quark parton distributions.

The unpolarized quark distributions in the nucleon are known
due to several fits \cite{mrs,fits} to the experimental data. 
The one of
the most recent fit is given by Martin, Roberts and Stirling (MRS) 
\cite{mrs} who used  existing experimental points in order to determine
quark and gluon distributions.

In this paper we would like to determine polarized quark parton distributions
starting from the MRS fit and using all existing data for proton, neutron
and deuteron spin asymmetries. We shall concentrate not only on a 
controversial low $x$ region but also on the behaviour of polarized  parton 
distributions at $x \rightarrow 1$. The phenomenological analysis of CERN and
SLAC data gives different results for $\Delta \Sigma$ (quark spin content
of the proton) and $\Delta s$ (strange quark polarization in the proton). 
Our results, as one will see later, prefers rather the numbers gotten from an 
analysis based on neutron E142 data.

Let us start with the formulas for unpolarized quark parton distributions
(at $ Q^{2}=4\, {\rm GeV^{2}}$) given by Martin, Roberts and Stirling. First of all
we shall consider their fit called $D'_{-}$ with very singular behaviour of
"sea" distribution at small $x$ values (which, however agrees with the results 
from HERA \cite{hera}). We have for the valence quarks distributions:

\begin{eqnarray}
u_{v}(x)+d_{v}(x)&=&1.422 x^{-0.58}(1-x)^{3.92}(1+2.59\sqrt x+4.21x),
\nonumber \\
d_{v}(x)&=&0.074 x^{-0.76}(1-x)^{4.67}(1+28.7\sqrt x+8.58x) ,
\end{eqnarray}

and for the "sea" ones:

\begin{eqnarray}
2\bar{u} (x)&=&0.4S(x)-\delta (x), \nonumber \\
2\bar{d} (x)&=&0.4S(x)+\delta (x),\\
2\bar{s} (x)&=&0.2S(x), \nonumber
\end{eqnarray}

where

\begin{equation}
S(x)=0.083x^{-1.5}(1-x)^{7.4}(1+8.57\sqrt x+15.8x),
\end{equation}

and

\begin{equation}
\delta (x)=0.164x^{-0.58}(1-x)^{7.4}.
\end{equation} 

The most important features of this fit are that  the 
valence "up" quark distribution dominates at $x \rightarrow 1$ and that 
the "sea" is not
SU(2) symmetric (having Lipatov type behaviour \cite{lip} at small $x$).
Because the unpolarized parton distributions are the sum of spin up and 
spin down distributions whereas the polarized ones are the difference of 
these functions our idea is just to split the numerical constants in 
formulas (1,3,4) in two parts in such a manner that we get positive 
defined distributions (which is not so easy to achieve). 
Our expressions for $\Delta q(x) = q^{+}(x)-q^{-}(x)$
($q(x) = q^{+}(x)+q^{-}(x)$) are:

\begin{eqnarray}
\Delta u_{v}(x)+\Delta d_{v}(x)&=&x^{-0.58}(1-x)^{3.92}(a_{1}+a_{2}\sqrt
x+a_{3}x), \nonumber \\
\Delta d_{v}(x)&=&x^{-0.76}(1-x)^{4.67}(b_{1}+b_{2}\sqrt x+b_{3}x), 
\nonumber \\
&& \\
\Delta S(x)&=&cx^{-0.58}(1-x)^{7.4}, \nonumber \\
\Delta \delta (x)&=&dx^{-0.58}(1-x)^{7.4}. \nonumber
\end{eqnarray}

We want to stress that we are using very simple way of parameterization
of $q^{+}$ and $q^{-}$ without introducing any additional powers of $x$ or 
$(1-x)$, so only the coefficients for polarized quark distributions are 
fitted.
In order to get the integrable function $\Delta S(x)$ we divide two first 
coefficients in eq.(3) into  equal parts.

We fit our formulas (with eight parameters) to the experimental data
on spin asymmetries, given by:

\begin{eqnarray}
A^{p}_{1}(x)&=&\frac{4\Delta u_{v}(x)+\Delta d_{v}(x)+2.2 
\Delta S(x)-3\Delta \delta (x)}{4u_{v}(x)+d_{v}(x)+2.2
S(x)-3\delta (x)},  \nonumber \\
&& \\
A^{n}_{1}(x)&=&\frac{\Delta u_{v}(x)+4\Delta d_{v}(x)+2.2 
\Delta S(x)+3\Delta \delta (x)}{u_{v}(x)+4d_{v}(x)+2.2
S(x)+3\delta (x)}. \nonumber
\end{eqnarray}

In this paper we assume that the spin asymmetries do not depend on $Q^{2}$ 
what is suggested by the experimental data \cite{d,n} and phenomenological 
analysis \cite{alt}. The unknown parameters in
eq.(5) are determined by making best fit to the measured spin
asymmetries for proton (SLAC-Yale, EMC, SMC) and neutron (E142). We get 
the following values (taking care of positivity for quark distributions):

\begin{equation}
\begin{array}{lll}
a_{1}=0.874,&a_{2}=5.023,&a_{3}=12.73,\\
b_{1}=0.074,&b_{2}=0.884,&b_{3}=0.649,\\
c=-0.556,&d=-0.004.&
\end{array}
\end{equation}

It is interesting to note that the fit shows no significant SU(2) symmetry
breaking for "sea" polarization ($d$ coefficient is close to zero).
The $\chi ^{2}$ is 20.1 for 34 degrees of freedom for such a fit. In figures 
1a, 1b and 2 
the comparison of our fit with the experimental points for proton and
neutron asymmetries is given. The {\em prediction} for the deuteron case 
(which adds 7.2 to $\chi ^{2}$  for 11 degrees of freedom) with $A^{d}_{1}$ 
given by the following expression:

\begin{equation}
A^{d}_{1}(x)=\frac{5\Delta u_{v}(x)+4.4\Delta S(x)}{5u_{v}(x)+4.4S(x)}
(1-\frac{3}{2}p_{D}),
\end{equation}

($p_{D}$ is D-state probability equal to $5.8\%$) is presented in
figure 3. We see that in this case the calculated curve at small $x$ lies 
above the experimental points. This is
because we get positive $A^{d}_{1}$ for all values of
Bjorken variable, whereas the data are negative in small $x$ region.
Of course we can make a fit with inclusion of SMC deuteron data. Than we 
get $\chi ^{2}=26.6$ (for 45 degrees of freedom) that is not much less
than for proton+neutron case and predicted deuteron asymmetry 
($\chi ^{2}=20.1+7.2=27.3$). Making the fit to proton+deuteron 
asymmetries we get $\chi ^{2}=22.2$ (37 degrees of freedom) which 
become 41.3 when one adds the result for $A^{n}_{1}$ ($\chi ^{2}=19.1$
for eight points).
Hence, it is possible to get a satisfactory deuteron asymmetry using 
proton+neutron data and is not when one tries to get the neutron
asymmetry from proton+deuteron data. This is our $\chi^2$ argument why in 
our fit we use SLAC neutron data omitting
CERN deuteron ones. As we will see below such preference is 
justified also when one analyzes an integrated quantities.

We can use our fit to determine the first moments of parton distributions.
We get e.g. for $I=\int^{1}_{0}g_{1}(x,Q^{2})\, dx$ at $Q^{2} =4\, {\rm GeV^{2}}$:

\begin{eqnarray}
I^{p}& = &\frac{4}{18}\Delta u+\frac{1}{18}\Delta d+\frac{1}{18}\Delta s 
= 0.178, \nonumber \\
&& \\
I^{n}& = &\frac{1}{18}\Delta u+\frac{4}{18}\Delta d+\frac{1}{18}\Delta s 
= -0.027. \nonumber
\end{eqnarray}

Other combinations (singlet and octet ones) of quark polarizations are:

\begin{eqnarray}
a_{0}& = &\Delta \Sigma = \Delta u+\Delta d+\Delta s = 0.50, \nonumber \\
a_{3}& = &\Delta u-\Delta d = 1.23,  \\
a_{8}& = &\Delta u+\Delta d-2\Delta s = 0.71, \nonumber
\end{eqnarray}
whereas $\Delta s = -0.07$. 

We would like to stress that these results are obtained from the fit 
without any constraints (e.g. for integrated quantities), 
which is not a case in other fits (see e.g. 
ref.\cite{stir}). We also do not use any information from neutron and 
hyperon $\beta$-decays.
The value of $a_{3}$ should be compared to 
$g_{A}=1.257\pm0.003$ \cite{pdg} lowered by several percent due to QCD
corrections (see e.g. ref.\cite{smc1}), i.e. to $a_{3}\simeq1.1$. The
agreement is not perfect, although satisfactory. Also the value of 
$I^{p}$, when compared with the last published number:
$I^{p}=0.136\pm0.011\pm0.011$ \cite{p}, is too high. This happens
because the structure function $F_{1}(x)$ given
by the MRS fit lies few percent higher (in considered region) than the 
one used in phenomenological
analysis by EMC and SMC. Because the ratio of unpolarized to polarized 
structure functions, i.e. the asymmetry is fitted very well the 
result for $g_{1}(x)$ is also overshooted. 
The second difference arises because the 
experimental groups use "Regge" type extrapolation in the $x\rightarrow 0$ 
region. For example, the latest SMC value for $I^{p}$ is gotten assuming
constant value for $g_{1}(x)$ in unmeasured small $x$ region. When we 
modify our distributions in such a manner 
(only for $x$ between 0 and 0.003) we get:

\begin{equation}
\begin{array}{lll}
I^{p}=0.168,&I^{n}=-0.038,&\Delta \Sigma=0.43,\\
a_{3}=1.23,&a_{8}=0.62,&\Delta s=-0.06,
\end{array}
\end{equation} 
the values which are not substantially different from those presented in 
eqs.(9,10). The second reason for using this modification (for small $x$)
is that MRS distribution for $u_{v}(x)$ is not positive defined for tiny
$x'$s ($x\sim10^{-7}$) and hence such blunder is also present in our 
polarized distributions for $u_{v}^{+}(x)$ and $u_{v}^{-}(x)$. 

If we fit the polarized parton distributions to all measured asymmetries
(on proton, neutron and deuteron targets) we get e.g. $a_{3}=1.50$ which 
is approximately $40\%$ to big. If one uses  
$D_{0}^{'}$ fit of Martin, Roberts and Stirling instead of $D_{-}^{'}$ 
($D_{0}^{'}$ is not so divergent at $x\sim 0$) we get the similar values 
for integrals 
$I^{p}$ and $I^{n}$, whereas $a_{3}=1.34$. Also the $\chi^{2}$ is worse in 
this case. Hence, we prefer our distributions fitted to asymmetries 
measured on proton (SLAC-Yale, EMC, SMC) and neutron (SLAC E142) targets
and having its roots in the $D_{-}^{'}$ fit.

Now, we would like to make some comments about the $x \rightarrow 1$ 
behaviour of valence quark distributions. Looking at the data points for 
proton spin asymmetry the value close to 1 at $x \sim 1$ is preferred, 
whereas for neutron and deuteron case values close to 0 seem to be 
natural (such observation is fragile due to the big experimental errors 
in this $x$ region). In our approach we can give predictions for the 
behaviour of polarized quark distributions and spin asymmetries in the 
$x \rightarrow 1$ limit. Let us assume that valence $u$ and $d$ quark 
polarized distributions behave at $x \rightarrow 1$ as:

\begin{eqnarray}
u_{v}^{\pm}& \rightarrow &a_{\pm}(1-x)^p+\ldots,
\nonumber \\
d_{v}^{\pm}& \rightarrow &b_{\pm}(1-x)^p+\ldots,
\end{eqnarray}
where $p$ is the smallest power of $(1-x)$ term. Than, we have in such
a limit:
\begin{equation}
\frac{F_{1}^{n}}{F_{1}^{p}} \rightarrow \frac{a_{+}-a_{-}+4(b_{+}-b_{-})}
{4(a_{+}+a_{-})+b_{+}+b_{-}}
\end{equation}
and:
\begin{eqnarray}
A^{p}_{1}(x)& \rightarrow & \frac{4(a_{+}-a_{-})+b_{+}-b_{-}}
{4(a_{+}+a_{-})+b_{+}+b_{-}} , \nonumber \\
A^{n}_{1}(x)& \rightarrow & \frac{a_{+}-a_{-}+4(b_{+}-b_{-})}
{a_{+}+a_{-}+4(b_{+}+b_{-})} .
\end{eqnarray}

In the case of SU(6) symmetry the flavour-spin part of nucleon 
wave function gives: $a_{+}:a_{-}:b_{+}:b_{-} = 5:1:1:2$. Hence, one gets 
well known results: $F_{1}^{n}/F_{1}^{p} \rightarrow 1/4$, whereas $A_{1}^{p}
\rightarrow 5/9$ and $A_{1}^{n} \rightarrow 0$. The counting rules for 
parton distributions at $x \rightarrow 1$ (see ref.\cite{brod}) yield:
$a_{-}/a_{+}=b_{-}/b_{+}=0$, so one has $A_{1}^{p},A_{1}^{n} 
\rightarrow 1$. Authors of ref.\cite{brod} assume in addition:
$a_{+}/b_{+}=5$ (as in the SU(6) symmetric case), hence they get $F_{1}^{n}/F_{1}^{p} 
\rightarrow 3/7$. In the MRS fit one has $b_{\pm}/a_{\pm}=0$ which 
leads to $F_{1}^{n}/F_{1}^{p} 
\rightarrow 1/4$. For the spin asymmetries  one gets 
$A_{1}^{p}=A_{1}^{n} \rightarrow (a_{+}-a_{-})/(a_{+}+a_{-}) 
\leq 1$. In our case we have $A_{1}^{p}=A_{1}^{n}=A_{1}^{d} \cong 0.77$.
The different limits (at $x \rightarrow 1$) for $A_{1}^{p}$ and 
$A_{1}^{n}$ one can get assuming that coefficients $b$ are not 
negligible in comparison to $a'$s (see eq.(14)). But then it is 
impossible to get the value for $F_{1}^{n}/F_{1}^{p}$ suggested
by the experimental data, namely 1/4 (see the figures in ref.\cite{stir}).

We have not considered gluon distributions in our fit (there is no need to 
take it into account dealing with functions at fixed $Q^{2}$). If, however 
we include explicit gluon terms into the 
asymmetries (in the way proposed in ref.\cite{ross}) we get a fit 
which is worse than the considered one.

Starting from the MRS fit \cite{mrs} to the unpolarized deep inelastic 
scattering  data we made a fit to proton and neutron spin asymmetries 
in order to obtain polarized quark parton distributions. We got 
$\Delta u=0.91$ ($\Delta u_{v}=1.04$), $\Delta d=-0.33$ ($\Delta d_{v}=-0.18$) 
and $\Delta s=-0.07$ for integrated quantities. At $x \rightarrow 1$ 
the asymmetries for nucleons point towards the value equal to 0.77.
With the improved behaviour at $x \rightarrow 0$ for the unpolarized
parton distributions and consistent data for spin asymmetries (mainly
for deuteron) our method of determination of quark polarized distributions 
looks promising.

\vspace{0.5cm}
{\bf Acknowledgements} \\
We would like to thank Jan Nassalski for fruitful discussions.

\newpage 

{\bf Figure captions}

\begin{itemize}
\item[ Figure 1a] The comparison of spin asymmetry on protons
(data points from SLAC-Yale, EMC and SMC experiments) 
with the curve gotten from our fit (eqs.(5,6,7)).
\item[ Figure 1b] The same as in figure 1a but with $x$ in logarithmic scale.
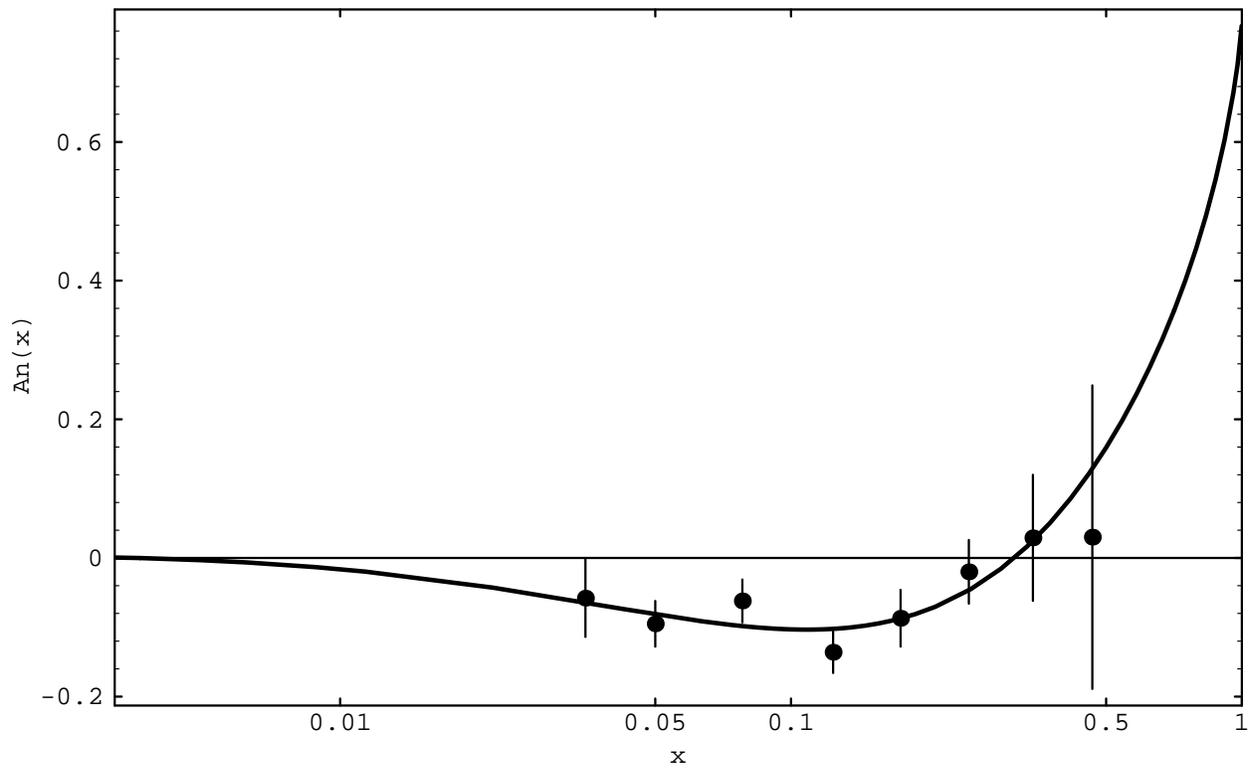
\item[Figure 2 \ ] The comparison of spin asymmetry on neutrons
(SLAC E142 data)
with the curve gotten from our fit (eqs.(5,6,7)).
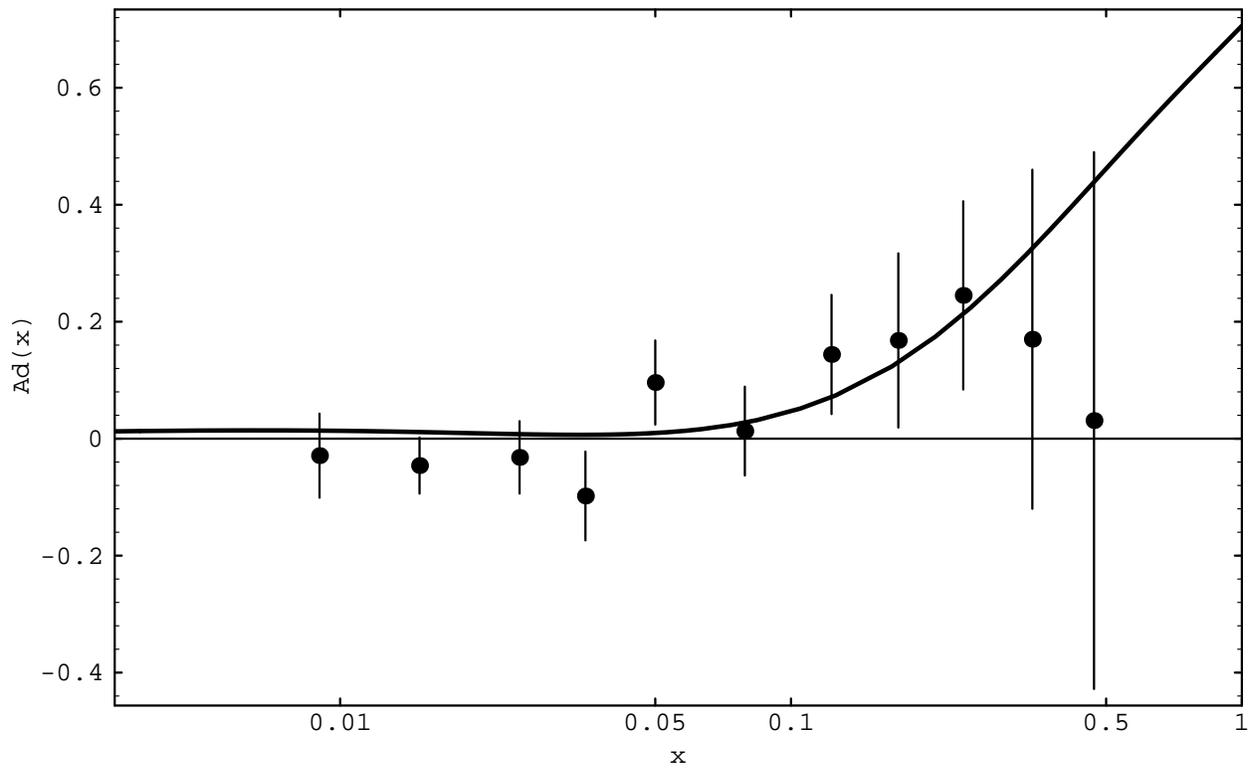
\item[Figure 3 \ ] Our prediction for deuteron asymmetry (NMC data).
\end{itemize}


\newpage
\begin{thebibliography}{99}
\bibitem{d} B.Adeva {\em et al.} (Spin Muon Collaboration), Phys.Lett. 
{\bf B 302}, 533 (1993);
\bibitem{p} D.Adams {\em et al.} (Spin Muon Collaboration), Phys.Lett. 
{\bf B 329}, 399 (1994); 
\bibitem{n} D.L.Anthony {\em et al.} (E142 Collaboration), Phys.Rev.Lett.  
{\bf 71}, 959 (1993);
\bibitem{pold} M.J.Alguard {\em et al.} (SLAC-Yale Collaboration), Phys.Rev.Lett. 
{\bf 37}, 1261 (1976); G.Baum {\em et al.}, Phys.Rev.Lett. {\bf 45},
2000 (1980); {\bf 51}, 1135 (1983);
\bibitem{emc} J.Ashman {\em et al.} (European Muon Collaboration), 
Phys.Lett. {\bf B 206}, 364 (1988); Nucl. Phys. {\bf B 328}, 1 (1989);
\bibitem{mrs} A.D.Martin, W.J.Stirling and R.G.Roberts, Phys.Rev. {\bf D 47},
867 (1993); Phys. Lett. {\bf B 306}, 145 (1993); {\em erratum} {\bf B 309},
492 (1993);
\bibitem{fits} J.F.Owens, Phys.Lett. {\bf B 266}, 126 (1991);
P.Chiapetta, G.Nardulli, Z.Phys. {\bf C 51}, 435 (1991); 
L.W.Whitlow {\em et al.}, Phys.Lett. {\bf B 282}, 475, (1992);
M.Gl\"{u}ck, E.Reya, A.Vogt, Z.Phys. {\bf C 53}, 127 (1992);
M.Gl\"{u}ck, E.Reya, A.Vogt, Phys.Lett. {\bf B 306}, 391 (1993);
J.Bolts {\em et al.}, Phys.Lett. {\bf B 304}, 159 (1993);
\bibitem{hera} I.Abt {\em et al.}, ($H_{1}$ Collaboration), Nucl.Phys. 
{\bf B 407}, 515 (1993); M.Derrick {\em et al.}, (ZEUS Collaboration), 
Phys.Lett. {\bf B 316}, 412 (1993);
\bibitem{lip} E.Akuraev, L.N.Lipatov, V.S.Fadin, Sov.Phys. JEPT {\bf
45}, 199 (1977); Ya.Balitsky, L.N.Lipatov, Sov.J.Nucl.Phys. {\bf 28},
822 (1978); I.Abt {\em et al.} ($H_1$ Collaboration), Phys.Lett. 
{\bf B 321}, 161 (1994);
\bibitem{alt} G.Altarelli, P.Nason, G.Ridolfi, Phys.Lett. {\bf B 320}, 152 
(1994);
\bibitem{stir} T.Gehrmann, W.J.Stirling, DPT/94/38 preprint (1994);
C.Bourrely, J.Soffer, CPT-94/P.3032 preprint (1994);
\bibitem{pdg} Particle Data Group, Phys.Rev. {\bf 45} (1992).
\bibitem{smc1} B.Adeva {\em et al.} (Spin Muon Collaboration), 
Phys.Lett. {\bf B 320}, 400 (1994);
\bibitem{brod} J.Brodsky, M.Burkardt, I.Schmidt, SLAC-PUB-6087 preprint,
(1994);
\bibitem{ross} G.Altarelli, G.G.Ross, Phys.Lett. {\bf B 212}, 391 (1988).
\end{thebibliography}
\end{document}